\documentclass{emulateapj}

\usepackage{natbib}

\slugcomment{To Appear in ApJ Letters}

\shorttitle{The SN 2011dh Progenitor Has Vanished}
\shortauthors{Van Dyk et al.}

\begin{document}

\title{The Progenitor of Supernova 2011dh Has Vanished}

\author{Schuyler D.~Van Dyk\altaffilmark{1},
  WeiKang Zheng\altaffilmark{2},
  Kelsey I.~Clubb\altaffilmark{2},
  Alexei V.~Filippenko\altaffilmark{2},
  S.~Bradley Cenko\altaffilmark{2,3},
 Nathan Smith\altaffilmark{4},
 Ori D.~Fox\altaffilmark{2},
 Patrick L.~Kelly\altaffilmark{2},
 Isaac Shivvers\altaffilmark{2},
 and Mohan Ganeshalingam\altaffilmark{5}
}

\altaffiltext{1}{Spitzer Science Center/Caltech, Mailcode 220-6,
  Pasadena CA 91125; email: vandyk@ipac.caltech.edu.}
\altaffiltext{2}{Department of Astronomy, University of California,
  Berkeley, CA 94720-3411.}
\altaffiltext{3}{Astrophysics Science Division, NASA Goddard Space Flight Center, Mail Code 661, Greenbelt, MD, 20771, USA}
\altaffiltext{4}{Steward Observatory, University of Arizona,
  Tucson, AZ 85720.}
\altaffiltext{5}{Lawrence Berkeley National Laboratory, Berkeley CA, 94720.}
  
 \begin{abstract}
We conducted {\sl Hubble Space Telescope\/} ({\sl HST})
Snapshot observations of the Type IIb Supernova (SN) 2011dh in M51 
at an age of $\sim 641$ days 
with the Wide Field Camera 3. We
find that the yellow supergiant star, clearly detected in pre-SN {\sl HST\/} images, has
disappeared, implying that this star was almost certainly the progenitor of the SN.
Interpretation of the early-time SN data which led to the inference of a compact nature for the 
progenitor, and to the expected survival of this yellow supergiant, is now clearly incorrect.
We also present ground-based $UBVRI$ light curves obtained with the Katzman Automatic Imaging 
Telescope (KAIT) at Lick Observatory up to SN age $\sim 70$ days. From the light-curve shape including 
the very late-time {\sl HST\/} data, and from recent interacting binary models for  
SN 2011dh, we estimate  that a putative surviving companion star to the now deceased yellow 
supergiant could be 
detectable by late 2013, especially in the 
ultraviolet. No obvious light echoes are detectable yet in the SN environment.
 \end{abstract}


\section{Introduction}\label{intro}

\bibpunct[;]{(}{)}{;}{a}{}{;} 
%\bibpunct[ ]{(}{)}{;}{a}{}{;}

Supernova (SN) 2011dh (also known as PTF11eon) in Messier 51 (M51; more precisely, M51a, or NGC 
5194) is a nearby example of the intermediate class of core-collapse supernovae (SNe), the Type IIb, 
existing between the hydrogen-rich SNe~II and the hydrogen-stripped SNe~Ib (see \citealt{filippenko97} for a review of SN classification). 
The progenitors of SNe~II, especially the Type II-Plateau SNe
(the most common core-collapse 
events), have been shown  to be red supergiants (RSGs) through direct progenitor 
identifications \citep[e.g.,][]{smartt04,fraser11,vandyk12a,vandyk12b}. 
The RSG progenitors of SNe~II-P appear to be consistent with expectations of single-star evolution 
models \citep[e.g.,][]{falk77,wheeler93,dessart11}.
The progenitors of both SNe~Ib and SNe~IIb, on the other hand, require the outer stellar envelope to
be substantially stripped away prior to explosion. One possible channel has long been thought 
to be massive interacting binary systems 
\citep[e.g.,][]{pod93,woosley94,maund04,claeys11}. SNe~IIb thus 
provide us with vital information regarding the evolutionary transition from a single massive red 
supergiant, with relatively low late-stage mass loss, to a star for which the mass loss before explosion 
must have been far more vigorous, potentially as a result of mass exchange with a companion. 
SNe~IIb constitute about 10--11\% of all core collapse SNe \citep{smith11,li11}, and their 
implied mass range appears to be consistent with binary evolution.
\citet{chevalier10} have further separated the already rare SNe~IIb into those whose
progenitors are compact (radius $R \approx 10^{11}$ cm) and those that are extended
($R \approx 10^{13}$ cm), based on emission from the shock-heated envelope and the radio
and X-ray properties of the SNe.
Hence, all nearby, well-studied cases of SNe~IIb are particularly valuable.

SN 2011dh was discovered independently by several amateur astronomers 
\citep[as summarized by][]{griga11} 
and by the Palomar Transient Factory (PTF) collaboration \citep{arcavi11} within $\sim 1$ day of
explosion, between May 31 and June 1.
Early photometry and spectra of the SN were presented by \citet{arcavi11}, who argued that the
data, when compared to analytical models of SN shock breakout, 
were {\em inconsistent\/} with an extended progenitor. 
Based on early-time radio
and X-ray data, \citet{soderberg12} also reasoned that the progenitor must have been compact.
Further radio observations were compiled by \citet{krauss12}, \citet{bietenholz12}, and 
\citet{horesh12}. Additional
X-ray data were collected and analyzed by \citet{horesh12}, \citet{campana12}, and \citet{sasaki12}.
\citeauthor{krauss12}~and \citeauthor{bietenholz12}~maintained that the progenitor was compact.
\citeauthor{horesh12}~concluded that the uncertainties in the modeling of the existing X-ray and 
radio data were larger than previously estimated, and therefore the inferred progenitor radius
is consistent both with being 
compact \citep[as found for the SN~IIb 2008ax;][]{pastorello08,taubenberger11,
chornock11} and extended
\citep[best exemplified by the SN~IIb 1993J in M81; e.g.,][]{richmond96,matheson00}.

Shortly after discovery of SN 2011dh, \citet{liavf11} identified
a possible progenitor star in archival {\it Hubble Space Telescope\/} 
({\it HST}) images of M51 obtained with the Advanced Camera for Surveys (ACS).
This identification was confirmed by both \citet{maund11} and 
\citet{vandyk11} by comparing ground-based adaptive optics images
of the SN with the archival {\it HST} ACS and
Wide-Field and Planetary Camera 2 (WFPC2) images.
The candidate progenitor had the bolometric luminosity ($L_{\rm bol} \approx 10^5\ {\rm L}_{\odot}$)
and effective temperature ($T_{\rm eff} \approx 6000$ K)
of a yellow supergiant (YSG). \citet{vandyk11},
following the conclusions of \citet{arcavi11} and \citet{soderberg12}, assumed that the 
progenitor star was compact and speculated that the YSG was the {\em companion\/} 
in an interacting binary system to the hotter, 
undetected progenitor. \citeauthor{vandyk11}~concluded that the YSG's initial mass was
$M_{\rm initial}=17$--$19\ {\rm M}_{\odot}$, suggested by the star's locus in the
Hertzsprung-Russell diagram (HRD) compared with the tracks of luminosity and effective
temperature for single massive-star theoretical evolutionary models. In contrast,
\citet{maund11} concluded that the YSG was, in fact, the 
likely progenitor star, with $M_{\rm initial}=13 \pm 3\ {\rm M}_{\odot}$, comparing its locus in the HRD 
only to the endpoint luminosities of the evolutionary tracks. \citet{murphy11} analyzed the stellar 
populations around the progenitor and found that, if the YSG had indeed vanished, then the 
progenitor likely had $M_{\rm initial} \approx 13\ {\rm M}_{\odot}$, in agreement with 
\citeauthor{maund11}

\citet{bersten12} subsequently performed hydrodynamical modeling of the existing light curves of
SN 2011dh. In particular, they showed that a more extended progenitor 
(with radius $R \approx 200\ {\rm R}_{\odot}$) was necessary to 
account for the post-shock-breakout thermal cooling and luminosity decline within the first 
$\sim 4$ days after explosion, indicated by the early $g'$-band data from \citet{arcavi11}.
This implies that the progenitor star at explosion was a supergiant with a low-mass H envelope 
($\sim 0.1\ {\rm M}_{\odot}$). \citeauthor{bersten12}~concluded from
their models that the progenitor's initial mass was consistent with 12--15 M$_{\odot}$.
The He core mass in the models was $\sim 4\ {\rm M}_{\odot}$, which, as 
\citeauthor{bersten12}~pointed out, is more consistent with the lower core mass expected from
an interacting binary model than with a single-star progenitor model.
However, the presence of a putative companion was entirely hidden by the light from 
the supergiant; both \citet{maund11} and
\citet{vandyk11} showed that the observed spectral energy distribution
of the supergiant could account for all of the flux in each {\sl HST\/} band, including the 
ultraviolet (UV) F336W band.

\citet{benvenuto13} continued exploring the binary scenario. For 
assumed initial masses of $16\ {\rm M}_{\odot}$ for the YSG primary (donor) star and 
$10\ {\rm M}_{\odot}$ for the secondary 
(a mass, and therefore luminosity, low enough to allow the secondary to remain 
undetected even in the pre-SN UV image), 
and an initial 
orbital period of 125 d, these authors found that, through three episodes of mass transfer, the primary
explodes with properties similar to those observed for the identified YSG and the secondary
remains near a hot zero-age main sequence (ZAMS) temperature and luminosity. The mass, 
temperature, and luminosity of
the secondary were all found to 
increase as the mass-transfer efficiency in the model interacting binary system was increased.

Ultimately, the key to determining the nature of the progenitor is to observe the SN at sufficiently
late times to reveal that the progenitor no longer is there. This sort of analysis has been conducted
in recent years to good effect by \citet[][ for SNe 1993J and 2003gd]{maund09}, 
\citet[][ for SN 2005gl]{galyam09}, 
\citet[][ in particular, for SNe 2004A, 2005cs, and 2006my]{maund13}, 
and \citet[][ for SN 2008bk]{vandyk13b}.
In this {\it Letter} we show that the YSG at the position of SN 2011dh 
has, in fact, disappeared (see also \citealt{vandyk13}, where we
initially announced this discovery). We also present multi-band photometry for 
the 
SN from day 5 to day 70 using the 0.76\,m Katzman Automatic Imaging Telescope \citep[KAIT;][]{filippenko01} 
at Lick Observatory. UT dates are used throughout (i.e., UTC, which is an approximation for UT1).

\section{Observations and Analysis}

\begin{figure}
\figurenum{1}
\plottwo{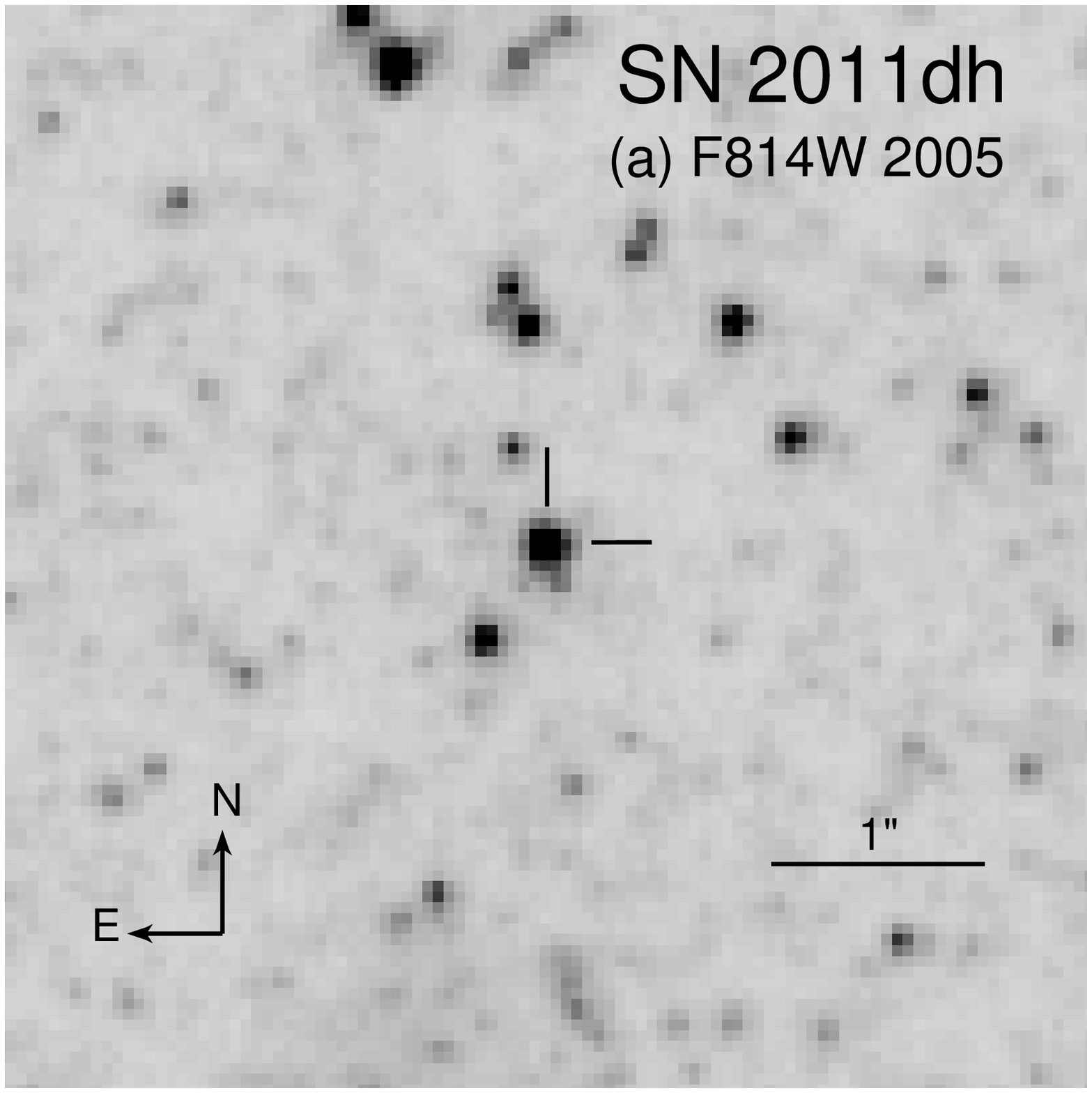}{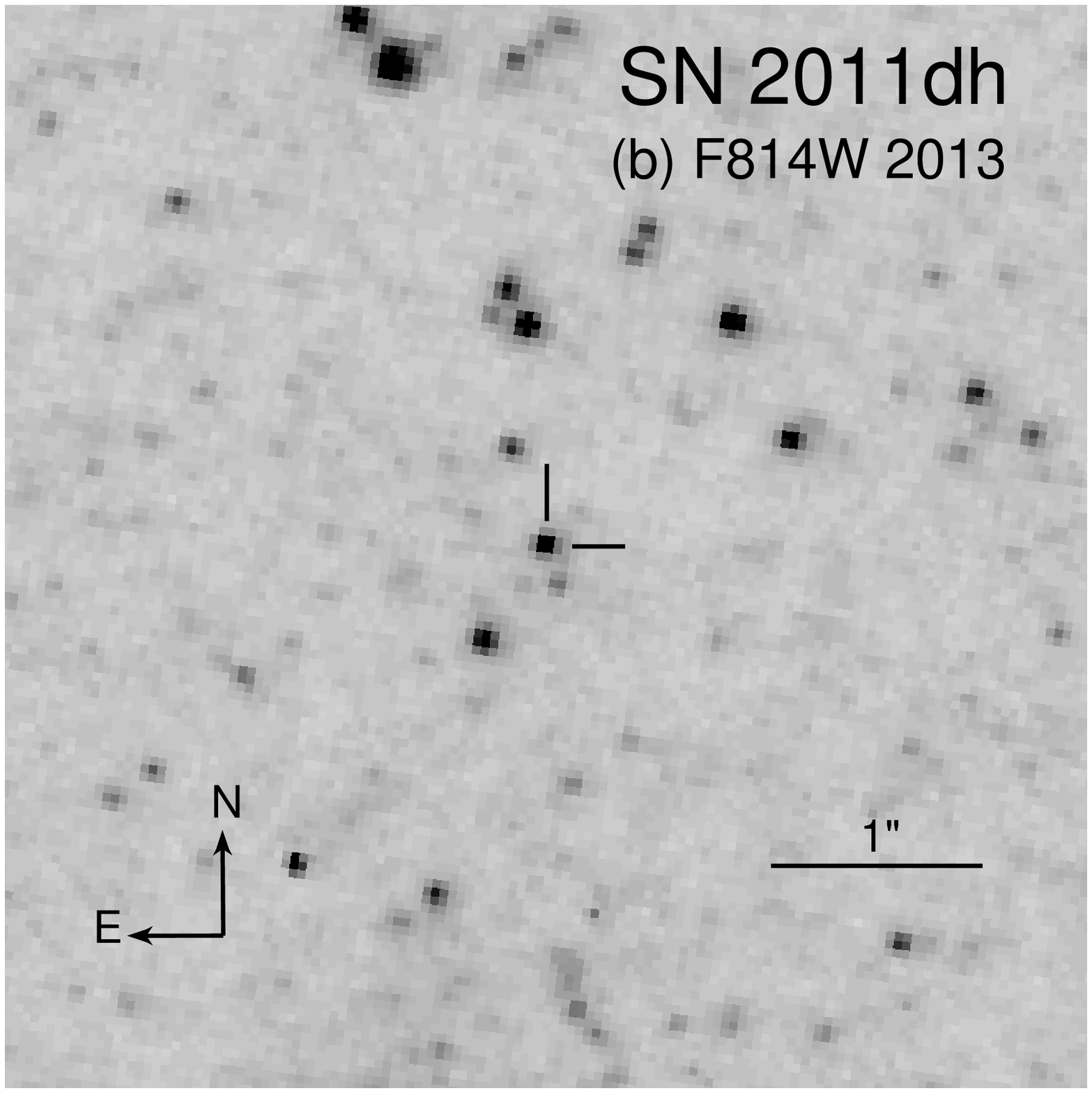}
\caption{(a) A portion of the archival {\sl HST} ACS F814W image of M51 from 2005; the
progenitor of SN 2011dh is indicated by {\it tickmarks}. 
(b) A portion of the {\sl HST} WFC3 F814W image from 2013, to the same scale and
orientation, and approximately the same contrast level; the SN is also indicated by {\it tickmarks}. 
North is up, and east is to the left.\label{figprog}}
\end{figure}

We observed SN 2011dh on 2013 March 2.44 (at age $\sim 641$ days) with {\sl HST\/}
using the Wide Field Camera~3 (WFC3) UVIS channel and filters F555W and F814W. 
These observations are part of our Cycle~20 Snapshot program GO-13029 (PI: A.~V.~Filippenko).
We display in Figure~\ref{figprog}b the SN as seen in F814W. 
For comparison we show in panel (a) 
the {\sl HST\/} ACS F814W image of the progenitor from 2005, 
with the same scale and orientation, and approximately the same contrast; 
the figure is similar to the one presented by \citet[][ their Fig. 1]{vandyk11}.

We extracted photometry from the WFC3 images using Dolphot v2.0 \citep{dolphin00}.
We present the {\sl HST\/} flight-system magnitudes for the SN in Table~\ref{hstphot}.
We have also remeasured photometry for the progenitor from the 2005 {\sl HST\/} ACS and WFPC2
images presented by 
\citet[][ also see \citeauthor{maund11}~\citeyear{maund11}]{vandyk11} using this version of Dolphot 
\citep[we had employed version 1.1 in][]{vandyk11}. We also used the pixel-based 
charge-transfer-efficiency-corrected ACS images available in the {\sl HST\/} archive.
The star is now measured to be somewhat brighter in all bands, although the differences at 
F336W 
and F814W are within the uncertainties (which are rather large at F336W and very small at F814W).
This new photometry is  presented in Table~\ref{hstphot}.
From the table one can see that the SN in 2013 March is 1.39 and 1.30 mag fainter in F555W and 
F814W, respectively, than was the YSG.

We have also observed the SN with KAIT in $UBVRI$ between 2011 June 3.25  (day $\sim 5$)
and August 8.18 (day $\sim 70$). 
We first subtracted away the light from the host galaxy using template images in $BVRI$ obtained at
the Lick Observatory Nickel 1\,m telescope on 2013 March 13 and in $U$ on 2013 May 8, 
when the SN 
was no longer detectable in these ground-based images. 
The templates were astrometrically 
registered, rescaled to match the pixel size of the KAIT images, and sky-background matched
before subtraction. Point-spread function (PSF) photometry was applied using the DAOPHOT \citep{stetson87} 
package from IDL Astronomy User's Library\footnote{http://idlastro.gsfc.nasa.gov/contents.html .}. 
The instrumental magnitudes and colors of the SN were transformed at $BVRI$ to the standard 
Johnson-Cousins system using the photometric sequence around the host galaxy presented by 
\citet[][ their Table A.5]{ergon13b}, specifically their stars P09-2, P09-3, P09-4, P09-A, and E13-1.
At $U$ we used only the two brightest of these stars, P09-2 and P09-4.
Uncertainties in the calibration of our photometry are 0.03 mag in $U$, 0.02 mag in $B$, and 
0.01 mag in $VRI$, and these have been added in quadrature with the measurement uncertainties.

We present the KAIT photometry in Table~\ref{kaitphot} and display the resulting light curves in 
Figure~\ref{figlc}. We also show for comparison the expected light-curve
decline rate (0.98 mag [100 day]$^{-1}$) powered by the radioactive decay of 
$^{56}$Co.

\citet{tsvetkov12} present $UBVRI$ photometry for the SN over the first 300 days.
\citet{marion13} and \citet{ergon13b} also provide UV-to-infrared photometry and 
spectroscopy of SN 2011dh over its first 34 and 100 days, respectively.
Our photometry compares quite well with that of \citeauthor{tsvetkov12}:  
differences are $\ll 0.1$ mag for $BVRI$, and our photometry in $U$ agrees very well
through maximum light, but is $\lesssim 0.4$ mag brighter post-maximum.
A similar favorable comparison also exists with the data from \citet{ergon13b}, although our $B$
photometry is $\sim 0.2$ mag brighter after day $\sim 38$, and our $U$ photometry is systematically
$\sim 0.2$ mag brighter. 
The remaining discrepancies in $U$ between our
photometry and that of \citeauthor{tsvetkov12}~and \citeauthor{ergon13b}~could result from
differences in application of a color term, which we did not apply.
More significant differences are generally found with the \citeauthor{marion13}~dataset: 
Our photometry is 0.8--1.0  
mag, 0.2--0.4 mag, 0.15--0.2 mag, and 0.2--0.3 mag brighter in $U$, $B$, $V$, and $I$, respectively, 
and $\lesssim 0.4$ mag brighter than the {\sl Swift\/} $U$ photometry (a discrepancy which could
result from differences in calibration); the best agreement  
is in $R$, with differences of $\ll 0.1$ mag. Overall, we cannot provide an explanation for the 
differences between our photometry and that presented by \citeauthor{marion13}

\section{Discussion and Conclusions}

It is evident from our recent {\sl HST\/} imaging
that the YSG has vanished, meaning that this is almost certainly the star that actually exploded, 
in agreement with the theoretical analyses by \citet{bersten12} and
\citet{benvenuto13}, and with the conclusions of \citet{maund11} and \citet{horesh12}.
A similar conclusion was reached by \citet{ergon13a,ergon13b} from their ground-based imaging.
Possible loci of the progenitor and its companion in the HRD are shown by \citet{benvenuto13}.
The model YSG  progenitor of SN 2011dh from \citet{bersten12} has radius 
$R \approx 200\ {\rm R}_{\odot}$, which is less extended than the progenitor
of the best-studied SN IIb 1993J: assuming the inferred absolute $V$ magnitude and 
K0 spectral type (effective temperature $\sim 4200$ K, with corresponding bolometric 
correction from \citeauthor{levesque05}~\citeyear{levesque05}) for the SN 1993J supergiant 
progenitor from \citet{vandyk02}, that star's radius was $\sim 580\ {\rm R}_{\odot}$. 
However, the
SN 2011dh progenitor is far more extended than, say, the radius inferred by 
\citet[][ $\sim 10^{11}$ cm]{chevalier10} for the SN~IIb 2008ax compact progenitor
(see, however, \citealt{horesh12} on the uncertainties in distinguishing between 
compact and extended progenitors based on radio data alone).
We speculate that the origin of compact versus extended progenitors of SNe~IIb, if this
distinction actually exists, may arise
in the initial conditions of the binary system (assuming the binary hypothesis applies for these SNe)
and the extent of mass exchange between the components. A more complete discussion of 
SN~IIb progenitors is beyond the scope of this {\it Letter}.

\begin{figure}
\figurenum{2}
\plotone{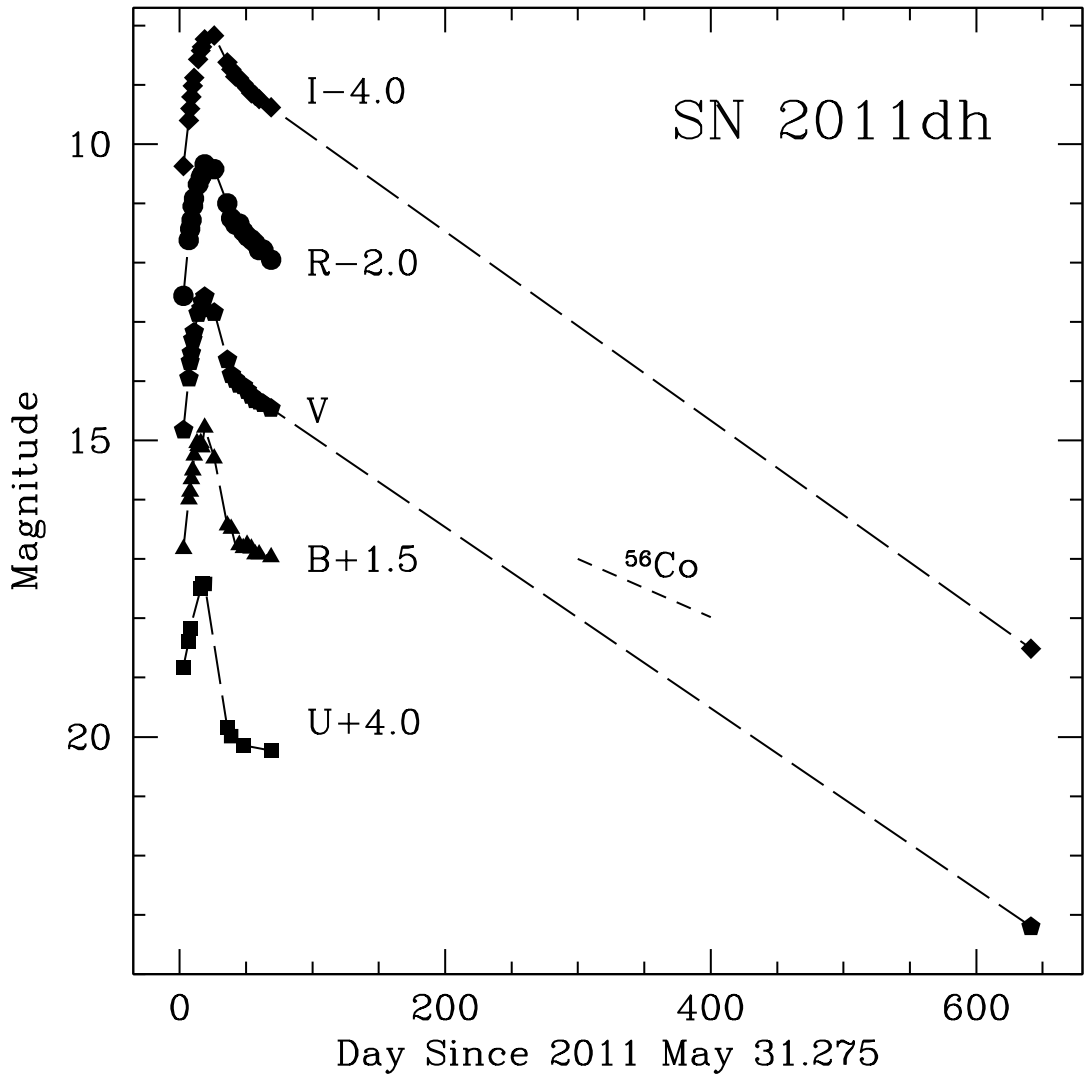}
\caption{Light curves of SN 2011dh, including data points from KAIT and {\sl HST}. Also shown
is the expected decline rate from the radioactive decay of $^{56}$Co (short-dashed line).
We interpolate between the last data points with KAIT to the {\sl HST\/} data points with the
long-dashed line. The {\sl HST\/} WFC3 F555W and F814W are assumed to be
${\sim}V$ and ${\sim}I$ bands. 
\label{figlc}}
\end{figure}

We interpolate in Figure~\ref{figlc} between 
the end of our photometric coverage of the SN with KAIT to
the {\sl HST\/} WFC3 data points. The behavior of the light curves shows that the SN light
declined over $\sim 641$ days more rapidly ($\sim 0.015$ mag day$^{-1}$ in $V$ and $I$)
than expected from $^{56}$Co decay, possibly as
a result of increasing transparency of the SN ejecta to the $\gamma$-rays emitted from the 
decay \citep[e.g.,][]{arnett89} or from dust formation.

We detect nebular emission lines from the SN in recent optical spectra;
\citet{shivvers13} present our analysis of them and 
discuss implications for the nature of the progenitor star.
These spectra show that light from the SN is still dominant at the time of the WFC3 observations.
The SN is therefore currently too bright for any binary companion of the progenitor to be detected.

However, we can estimate the earliest date at which the companion, if it exists, could become visible.
If we adopt the mass-transfer efficiency of 0.25 from 
\citet[][ their Table 3]{benvenuto13}, the secondary has $T_{\rm eff}\approx 30000$ K, 
$L_{\rm bol}=10^{4.13}\ {\rm L}_{\odot}$, and surface gravity $\log g \approx 4.3$ 
(essentially that of a late O- or early B-type supergiant). Assuming a
bolometric correction for a supergiant at this temperature from \citet{flower96}, 
an M51 distance
of 8.4 Mpc from \citet{vinko12}, and an extinction $A_V < 0.25$ mag
(adopting the Galactic foreground contribution from \citealt{schlafly11} and
the upper limit on the host-extinction contribution from \citealt{arcavi11},
with $R_V=3.1$), we find that the secondary's brightness would be $V \lesssim 27.2$ mag. 
(We note that \citealt{bersten12} also estimate that the progenitor's companion 
should have a visual magnitude of $\sim 26$--27.)
Given the current decline rate of the SN, we estimate that a surviving 
companion might
become visible at day $\sim 900$ --- that is, as early as 2013 mid-November.
Furthermore, given the expected high effective temperature of the surviving companion, a search
for this star is optimized in the UV, since the star's light should dominate at these wavelengths.
If the SN in the $U$ band has been declining at the same rate as in $V$ 
and $I$, then by day $\sim 900$ the SN should have become quite faint, at $U \approx 29$ mag.
(One caveat here is that the post-maximum $U$-band light curves shown by
\citealt{tsvetkov12} and especially \citealt{ergon13b}
appear to flatten out up to day $\sim 100$.)
Also, from \citet{benvenuto13}, if the mass-transfer efficiency (essentially a free parameter) 
is higher, detection would be 
possible sooner, since the star should be hotter and more luminous; 
however, if it is lower, we may have to wait longer to detect the star.

Finally, we note that no obvious, large-scale light echo has yet appeared in the 2013 {\sl HST\/} 
images. Assuming the VEGAMAG zero point for 
WFC3/UVIS\footnote{http://www.stsci.edu/hst/wfc3/phot{\textunderscore}zp{\textunderscore}lbn .} 
and a plate scale of
$0{\farcs}04$ pixel$^{-1}$, we place a $3 \sigma$ upper limit on the surface brightness 
of an extended echo at F555W of $\gtrsim 21.6$ mag arcsec$^{-2}$.
(We also cannot rule out that a more compact echo, within the image PSF, is contributing to 
the observed SN flux.)
However, the extended echoes around the SN~IIb 1993J
were not detected until $\sim 8$ yr after explosion 
\citep[][ echoes were not detectable $\sim 2$ yr post-explosion]{sugerman02,liu03}.
So, it is possible we may still see one or more echoes emerge 
in future {\sl HST\/} images of SN 2011dh, particularly in the UV and blue bands.

\acknowledgments 

This work is based in part on observations made with the NASA/ESA
{\it Hubble Space Telescope}, obtained from the Data Archive at the
Space Telescope Science Institute (STScI), which is operated by the
Association of Universities for Research in Astronomy (AURA), Inc.,
under NASA contract NAS 05-26555.   KAIT and its
ongoing research were made possible by donations from Sun
Microsystems, Inc., the Hewlett-Packard Company, AutoScope
Corporation, Lick Observatory, the NSF, the University of California,
the Sylvia \& Jim Katzman Foundation, and the TABASGO Foundation. 
Support for
this research was provided by NASA through grants AR-12623 and GO-13029 
from STScI.
A.V.F. and
his group at UC Berkeley also wish to acknowledge generous support from
Gary and Cynthia Bengier, the Richard and Rhoda Goldman Fund, the
Christopher R. Redlich Fund, the TABASGO Foundation, and NSF 
grant AST-1211916.  

{\it Facilities:} \facility{HST(ACS)}, \facility{HST(WFC3)}, \facility{Lick:KAIT}.

%\clearpage

\begin{deluxetable}{ccccccccc}
%\rotate
\tablewidth{7.4truein}
\tablecolumns{9}
\tablecaption{{\sl HST\/} Photometry at the Position of SN 2011dh\tablenotemark{a}\label{hstphot}}
\tablehead{
\colhead{Epoch} & \colhead{F336W} & 
\colhead{F435W} & \colhead{$B$} & 
\colhead{F555W} & \colhead{$V$} & 
\colhead{F658N} & \colhead{F814W} & \colhead{$I$} \\
\colhead{} & \colhead{(mag)} & \colhead{(mag)} & \colhead{(mag)}
& \colhead{(mag)} & \colhead{(mag)}
& \colhead{(mag)} & \colhead{(mag)}
& \colhead{(mag)}}
\startdata
2005 & 23.380(283) & 22.368(005) & 22.380 & 21.813(006) & 21.759 & 21.166(017) & 21.211(005) & 
21.203\\
2013 & \nodata & \nodata & \nodata & 23.198(019) & \nodata & \nodata & 22.507(022) & \nodata \\
\enddata
\tablenotetext{a}{Uncertainties $(1\sigma)$
are given in parentheses as millimagnitudes.}
\end{deluxetable}

%\clearpage

\begin{deluxetable}{cccccccc}
\tablewidth{5.4truein}
\tablecolumns{8}
\tablecaption{KAIT Photometry of SN 2011dh\tablenotemark{a}\label{kaitphot}}
\tablehead{
\colhead{JD$-$2,400,000} &
\colhead{$U$} & 
 \colhead{$B$} & 
\colhead{$V$} & 
\colhead{$R$} &
\colhead{$I$} \\
\colhead{} & \colhead{(mag)} & \colhead{(mag)}
& \colhead{(mag)} & \colhead{(mag)}
& \colhead{(mag)}}
\startdata
55715.75 & 14.82(08) & 15.33(08) & 14.82(05) & 14.56(05) & 14.37(03) \\
55719.77 & 14.39(07) & 14.49(08) & 13.95(06) & 13.62(06) & 13.60(05) \\
55720.73 & 14.17(07) & 14.36(09) & 13.67(05) & 13.43(05) & 13.40(04) \\
55721.70 & \nodata &     14.15(09) & 13.52(08) & 13.28(06) & 13.20(03) \\
55722.72 & \nodata &     14.00(09) & 13.30(05) & 13.05(06) & 13.02(04) \\
55723.73 & \nodata &     13.75(07) & 13.17(05) & 12.92(04) & 12.88(04) \\
55725.85 & \nodata &     13.54(06) & \nodata & \nodata & \nodata \\
55726.72 & \nodata &     13.59(07) & 12.86(04) & 12.68(04) & 12.57(04) \\
55728.72 & 13.50(13) & 13.54(08) & 12.78(05) & 12.56(06) & 12.42(03) \\
55729.73 & 13.42(07) & 13.61(09) & 12.66(03) & 12.51(06) & 12.36(02) \\
55731.73 & 13.42(06) & 13.28(02) & 12.57(02) & 12.34(02) & 12.23(02) \\
55738.72 & \nodata &     13.80(04) & 12.84(04) & 12.42(03) & 12.17(03) \\
55748.73 & 15.84(09) & 14.93(07) & 13.64(03) & 13.00(03) & 12.62(03) \\
55751.70 & 15.99(09) & 14.99(06) & 13.89(09) & 13.25(08) & 12.74(05) \\
55754.72 & \nodata &     \nodata     & 13.97(06) & 13.36(08) & 12.86(05) \\
55757.70 & \nodata &     15.26(05) & 14.05(06) & 13.34(03) & 12.89(02) \\
55760.72 & 16.14(08) & 15.31(06) & 14.08(03) & 13.47(05) & 12.98(02) \\
55763.70 & \nodata &     15.26(05) & 14.16(04) & 13.55(05) & 13.05(04) \\
55766.72 & \nodata &     15.32(04) & 14.24(05) & 13.61(05) & 13.14(04) \\
55769.72 & \nodata &     15.42(05) & 14.31(04) & 13.67(04) & 13.18(04) \\
55772.71 & \nodata &     15.42(05) & 14.34(04) & 13.79(06) & 13.24(06) \\
55775.71 & \nodata &         \nodata & 14.38(03) & 13.79(04) &  \nodata \\
55781.68 & 16.23(10) & 15.47(05) & 14.46(05) & 13.95(06) & 13.38(04) \\
\enddata
\tablenotetext{a}{Uncertainties $(1\sigma)$
are given in parentheses as hundredths of a magnitude.}
\end{deluxetable}

%\clearpage

%\clearpage

\end{document}